\documentstyle[aps]{revtex}

\begin{document}
\draft
\title{The hysteresis loop area of the Ising model}
\author{Han Zhu$^{1,\thanks{%
Present address: Department of Physics, Princeton University, Princeton, NJ
08544, USA}}$, Shuai Dong$^1$ and J. -M. Liu$^{1,2,\thanks{%
Author to whom correspondence should be addressed. Email address:
liujm@nju.edu.cn}}$}
\address{$^1$Laboratory of Solid State Microstructures, Nanjing University, Nanjing
210093, \\
$^2$International Center for Materials physics, Chinese Academy of Sciences,
Shenyang, China}
\maketitle

\begin{abstract}
The hysteresis of the Ising model in a spatially homogeneous AC field is
studied using both mean-field calculations and two-dimensional Monte Carlo
simulations. The frequency dispersion and the temperature dependence of the
hysteresis loop area are studied in relation to the dynamic symmetry loss.
The dynamic mechanisms may be different when the hysteresis loops are
symmetric or asymmetric, and they can lead to a double-peak frequency
dispersion and qualitatively different temperature dependence.
\end{abstract}


When a cooperative many-body system, such as a magnet, is placed in an
oscillating external perturbation (such as a magnetic field), it may show
also oscillating dynamic response. This response usually lags in time,
creating a hysteresis loop with nonzero area. This phenomenon exists widely
in, e.g., magnetic systems and ferroelectric systems\cite{ferrorev}, and has
been arousing great interest for its important technical application and
intriguing physics\cite{rev1,rev2,rev3}.

The recent theoretical\cite
{heisenberg,heisenberg2,LG,LJMferro,LJMphi,RG,sides,absence,Atemp} and
experimental\cite{expeA} studies on the hysteresis (also see Ref. \cite{rev3}
and references therein) focus on two topics: the dynamic symmetry breaking,
and the area of the hysteresis loop. The first phenomenon is due to the
competing time scales in such nonequilibrium systems\cite{rev3}: The
hysteresis loop loses its symmetry when the time period of the oscillating
external perturbation becomes much smaller than the typical relaxation time
of the system. On the other hand, the interesting variance of the hysteresis
loop area with such parameters as temperature and oscillation frequency can
also be attributed to the time scale competition. For example, in the
frequency dispersion of the loop area of the Ising model, the frequency $%
\omega _0/2\pi $ giving the maximal area corresponds roughly to the point
where a resonance occurs. When an Ising system is placed in an AC field, the
dynamics may consist of domain nucleation and/or domain growth\cite{rev2}.
The nucleation rate of new domains can be predicted by a characteristic time 
$\tau _n$, and the domain growth rate is also linked with a characteristic
time $\tau _g$. The resonance occurs when the time period of the external
perturbation is comparable to either one of these time scales or a
combination of them. As is shown below, the details of this dynamic time
scale competition necessarily rely on the dynamic phase of the system.

In the present work, we hope to help clarify the relationship of the two
mentioned-above topics in the framework of the Ising model, with mean-field
(MF) calculations and two-dimensional Monte-Carlo (MC) simulations. (The
frequency range that receives the most attention here is within the
discussion of the previous works using the same methods\cite{rev3}.) When
the loops are symmetric, the system dynamics is controlled by a domain
nucleation-and-growth mechanism. When the loops are asymmetric (especially
when the magnetization is well above or below zero), throughout the system
evolution we can observe most spins being in the same direction. The
dynamics of the remaining spins in the opposite direction may be described
mainly by the domain nucleation mechanism. In the following, we shall see
that the variance of the loop area with frequency, field amplitude and
temperature strongly depends on the dynamic mechanism, which is determined
by the loop symmetry.

{\it The model.} Before the results are presented, we first describe the
model. (1) The evolution of the magnetization $M$ in the mean-field Ising
model is determined by the following equation\cite{rev3}, 
\begin{equation}
\frac{dM}{dt}=-M+\tanh \left( \frac{M+H\left( t\right) }T\right) .
\label{MFequation}
\end{equation}
(2) In the MC simulation, the Hamiltonian of the two-dimensional Ising model
in a spatially homogeneous field $H\left( t\right) $ can be written as 
\begin{equation}
{\cal H}\left( \left\{ \sigma _i\right\} ;t\right) =-J\sum_{\left\langle
i,j\right\rangle }\sigma _i\sigma _j-H\left( t\right) \sum_i\sigma _i.
\end{equation}
The magnetization is obtained as $M=\sum_i\left\langle \sigma
_i\right\rangle $. The MC simulation goes as the following: The coupling
constant $J$ and the Boltzmann constant $k_B$ are both taken as $1$. On a
two-dimensional $N\times N$ (in the present work, $100\times 100$) lattice
with periodic boundary condition, at each time step a spin $\sigma _i$ is
randomly chosen and the probability that it is flipped is\cite{zjy} 
\begin{equation}
W\left( \sigma _i\rightarrow \hat{\sigma}_i\right) =\frac 1Q\exp \left[ -%
\frac 1T{\cal H}\left( \left\{ \hat{\sigma}_i,\sigma _{j\neq i}\right\}
;t\right) \right] ,
\end{equation}
where $\hat{\sigma}_i=\pm \sigma _i$ and $Q$ is the normalization factor.
Then in the next time step the field is updated and another spin is picked
at random. A MC step consists of $N\times N$ such unit procedures. The unit
time is chosen to be one MC step, with the time resolution being $1/N^2$.
The initial state is always $80\%$ randomly chosen spins up (MC simulation)
or $M=1$ (MF calculations). In both studies, the system evolves into
equilibrium after a zero-field relaxation. Then, an AC field, $H\left(
t\right) =H_0\sin \omega t$, is applied. The measurement of the hysteresis
loop always begins after a number of introductory cycles, and the symmetry
of the loop can be characterized by the order parameter $Q=\int_0^{2\pi
/\omega }Mdt$.

The phase diagram of the Ising model in an oscillating field has been
extensively studied\cite{rev3,absence,MFfirst,Ising95}. With regards to the
area scaling of the hysteresis loops of the Ising model, there are already
MC and MF results, and a detailed review can be found in Ref. \cite{rev3}.
Generally speaking, the area can be written as $A=A_{sta}+A_{kin}$, where $%
A_{sta}$ is the possibly nonzero static contribution existing even in the
quasistatic limit, and $A_{kin}$ is the kinetic contribution assuming the
form of 
\begin{equation}
A_{kin}\sim H_0^\alpha T^{-\beta }g\left[ \widetilde{\omega }\left( \omega
,H_0,T\right) \right].  \label{Apre}
\end{equation}
Here $g\left( \widetilde{\omega }\right) $ has been believed to be a
single-peak function, and $\widetilde{\omega }\left( \omega ,H_0,T\right) $
has been supposed to have the form of $\omega /\left( H_0^\gamma T^\delta
\right) $.\footnote{%
An exception is that, in the MF\ Ising model, with temperature $T$ above $%
T_c=1$, and $H_0$ very small, it is analytically obtained $A\sim
H_0^2T^{-1}\omega /\left( \varepsilon ^2+\omega ^2\right) $, where $%
\varepsilon =\left( T-1\right) /T$.\cite{rev3}} There have been extensive
theoretical efforts to determine the exact form of $g\left( \widetilde{%
\omega }\right) $, and the values and the physical meaning of the exponents.
As we shall see below, the dynamic phase transition may lead to a
double-peak frequency dispersion and a piecewise analytic function of
temperature dependence.

{\it The frequency dispersion of the hysteresis loop area. }Fig. 1 (a) shows
a typical result of the two-dimensional MC simulations, at temperature $%
T=1.0<T_c$. (When $T<T_c$, a dynamic symmetry loss can be observed as $H_0$
decreases from $4$ to $0$.\cite{rev3}) As is clearly shown in the curve with 
$H_0<2$, two distinct peaks can be observed. We name the left one as peak I
and the right one as peak II. It is found that peak I occurs in the range
where the hysteresis loops are symmetric and peak II is in the range of
asymmetric loops. As the frequency is increased from peak I to peak II, a
dynamic symmetry loss occurs.

Fig. 1 (b) shows a typical result of the MF calculations, at temperature $%
T=0.5<T_c$. Similarly, a double-peak function can be observed. However,
there is a very important difference. Given relatively small values of $H_0$%
, it is possible that there is only peak II in the MF calculation. This is
because, the equilibrium magnetization can be obtained by solving the
following equation, 
\begin{equation}
-M+\tanh \left( \frac{M+H}T\right) =0.  \label{stable}
\end{equation}
With $H$ small enough and $T<T_c=1$, there can be two stable solutions to
Eq. (\ref{stable}), corresponding to two values of stable equilibrium
magnetization (a positive one and a negative one). This means that the
hysteresis loops can be asymmetric even in the quasistatic limit. By
contrast, according to Ref. \cite{rev3}, in MC simulation the hysteresis
loop is always symmetric in the quasistatic limit, as a result of
fluctuation.

In the above discussions we provide evidence of two peaks I and II, which
correspond to the resonance of symmetric and asymmetric hysteresis loops
respectively. In the following we give a general explanation of the physics
origin of this observation. The existence of two peaks clearly indicate two
time scales, corresponding to two different dynamic mechanisms. When the
loops are symmetric, both the initial domain nucleation and the late stage
domain growth are at work. As suggested by Liu {\it et al.}\cite
{LJMferro,LJMphi}, the observation of peak I means that a third time scale $%
\tau _{\text{I}}$ can be defined as a combination of $\tau _n$ and $\tau _g$%
, and the resonance occurs as $2\pi /\omega _{\text{I}}\sim \tau _{\text{I}}$%
. Since for the asymmetric loops the magnetization can be always well above
or below zero, at any time during the system evolution we can observe most
spins having the same direction. The late stage domain growth is relatively
inhibited, and thus, the time scale $\tau _{\text{II}}$ corresponding to
peak II shall be mainly determined by $\tau _n$.

Compared with peak II, the resonance at peak I has been relatively well
studied in the previous works. Some illustrations of the resonance at peak I
can be found in, e.g., Ref. \cite{sides,LJMphi}. In the following we focus
on peak II, as illustrated in Fig. 2 (a) and (b). Actually, the existence of
peak II can be easily understood in the MF\ Ising model. As is mentioned
above, when $H$ is small enough, there can be two stable solutions of Eq. (%
\ref{stable}) . Thus, with $\omega \rightarrow 0$, the hysteresis loop
reduces to a curve (the stable solution of Eq. (\ref{stable}), as shown in
Fig. 2 (b). In the other limit, with $\omega \rightarrow \infty $, the
system cannot respond to the external field, and the hysteresis loop becomes
a horizontal line. Thus, it is straightforward to predict the existence of a
peak in the intermediate region, where the phase lag between the external
field and the system response creates a hysteresis loop with nonzero area.
Note that the observation of peak II requires temperatures lower than the
static critical point and small enough values of $H_0$. When $H_0\ll T\ll 1$%
, we can give a quantitative description of the system behavior by solving
the MF equation (\ref{MFequation}) analytically. We suppose $1-M\rightarrow
0 $, and obtain 
\[
\frac{dM}{dt}=1-M-2\exp \left( -\frac 2T\right) \left( 1-\frac 2TH_0\sin
\left( \omega t\right) \right) . 
\]
This equation can be exactly solved, and the loop area is obtained as 
\begin{eqnarray}
A &=&-\int_0^{2\pi /\omega }Md\left( H_0\sin \omega t\right)  \nonumber \\
&=&4\pi \left[ \frac 1T\exp \left( -\frac 2T\right) \right] H_0^2\frac \omega
{\omega ^2+1}.  \label{A}
\end{eqnarray}

In the following we report some important differences and similarities of
peak I and II, as summarized from Eq. (\ref{A}) and the numerical results in
Fig. 1. Differences: (1) The temperature dependence of the height of peak
II\ in the MF Ising model is obtained in Eq. (\ref{A}), and is different
from the previous $T^{-1/2}$ prediction of peak I\cite{Ising95}. (2) In the
MF Ising model, the maximal area $A_{\max }^{\text{II}}$ at peak II grows
with $H_0$ as $H_0^2$, while the maximal area $A_{\max }^{\text{I}}$ at peak
I has been predicted to grow with $H_0$ linearly in the previous studies\cite
{Ising95}. This difference is explained by observing the variance of the
loop shape with $H_0$, as illustrated in Fig. 3(a) and (b): At peak I, only
the width of the loop increases with $H_0$ (that is why $A_{\max }^{\text{I}%
} $ grows linearly with $H_0$), while at peak II, the loop is expanding in
two directions with $A_{\max }^{\text{II}}$ growing as $H_0^2$.

Similarities: (1) In both MF calculations and MC simulations, it is found
that, as $\omega \rightarrow \infty $, the area decays as $\omega ^{-1}$.
(With respect to peak I, this is in accordance with the previous MF result
and the work of Rao {\it et al.}\cite{phi1,phi2} on the $\left( \Phi
^2\right) ^2$ and $\left( \Phi ^2\right) ^3$ model, but not the previous MC
result of the Ising model\cite{rev3,Ising95}, which indicates an
exponentially decaying function of $g\left[ \widetilde{\omega }\left( \omega
,H_0,T\right) \right]$ in Eq. (\ref{Apre}).) (2) Independent of $H_0$ and $T$%
, peak II\ is always observed to be at (or very close to) $\omega_{\text{II}%
}=1$. With regards to peak I, in both MC simulations and MF calculations we
find that as , also approaches $1$, which is different from the previous
predictions. According to Ref. \cite{rev3,Ising95}, the $\omega_{\text{I}}$
will also tend to infinity as $H_0\rightarrow \infty $. But it is not what
we observe in Fig. 1, which shows that, as the field already far exceeds the
spin-spin interaction, the time scale of the system is no longer sensitive
to the value of $H_0$.

In the following we turn to study{\it \ the temperature dependence of the
loop area} with fixed field amplitude and frequency\cite{Ising95,Atemp}.
Here our motivation is quite similar to that of the above discussions of the
frequency dispersion. When $H_0<4$ (MC) or $H_0<1$ (MF), a dynamic symmetry
loss can be observed as $T$ decreases (as can be predicted from the phase
diagram previously obtained\cite{rev3,absence,MFfirst,Ising95}). Thus, a
simple scaling function is not likely to exist for the loop area, since
there are different dynamic mechanisms of the symmetric and asymmetric
loops, and different time scale competition. This is supported by the MF and
MC results.

A typical MF result is shown in Fig. 4 (a). The order parameter $\left|
Q\right| >0$ for lower temperature and $\left| Q\right| =0$ for higher
temperature, and at the dynamic critical point, a dynamic symmetry loss
occurs. In Ref. \cite{Atemp}, using the same methods as the present work,
Acharyya found that the area becomes maximum above the dynamic transition
point. Here, it is clear that the temperature dependence of the loop area
assumes different functions for the $\left| Q\right| >0$ states and $\left|
Q\right| =0$ states. These different functions are separated by the dynamic
critical point and the first order derivative, $\partial A/\partial T$, is
not continuous at the dynamic critical point. Thus, the temperature
dependence is a piecewise analytic function. A typical MC result is shown in
Fig. 4 (b), and it is roughly similar to the MF result. Although we do not
observe a notable discontinuity of the first order derivative, $\partial
A/\partial T$, it is very obvious that the second order derivative, $%
\partial ^2A/\partial T^2$, changes sign at the dynamic critical point. When 
$H_0>4$ (MC) or $H_0>1$ (MF), the field amplitude exceeds the spin-spin
interaction, and the hysteresis loops are always symmetric\cite
{rev3,absence,MFfirst,Ising95}. In this case, we observe that the loop area
decreases monotonically as the temperature grows.

To summarize, the hysteresis of the Ising model in an AC field, $H\left(
t\right) =H_0\sin \left( \omega t\right) $ is studied using both mean-field
calculations and two-dimensional Monte Carlo simulations. The frequency
dispersion and the temperature dependence of the loop area are studied in
relation to the dynamic symmetry loss. The dynamic mechanisms are different
when the hysteresis loops are symmetric or asymmetric. For symmetric loops,
the dynamics is a combined domain nucleation-and-growth process. By
contrast, for asymmetric loops well above or below the $M=0$ line, the
dynamics may be mainly domain nucleation. This framework is part of basic
current knowledge of hysteresis phenomena, and the observed frequency and
temperature dependence of the loop area is consistent with it. Double peaks
can be observed in the frequency dispersion, and the temperature dependence
is possibly a piecewise analytic function. Interestingly, the shift of the
dynamic mechanism with the symmetry loss is also found in the mean-field
calculation, and some striking similarities are observed (for example, the
same position of peak II). Although the present work deals with a model spin
system, the topics discussed have general meaning. Surely some quantitative
details like the position of the peaks rely on the model setting, but we
believe that the physics of the conclusions is not limited to the specific
system studied here, and can be predicted for more general systems.

We thank Hao Yu for helpful discussions. This work is supported by the
Natural Science Foundation of China (50332 020, 10021001) and National Key
Projects for Basic Research of China (2002CB61 3303).

\null\vskip0.2cm

Fig.1 The frequency dispersion of the loop area $A\left( \omega \right) $.
(a) The MC results with AC field at temperature $T=1.0$; (b) The MF results
with AC field at $T=0.5$. The dashed line, corresponding to $\omega^{-1}$,
serves as a guide to eye.

Fig.2 The illustrations of peak II with (a) MC simulations at $T=1.0$ and
(b) MF calculations at $T=0.5$.

Fig.3 (a) Some typical hysteresis loops at peak I, with $H_0=3,6,...,30$
from the innermost to the outermost loop. (b) Some typical hysteresis loops
at peak II, with $H_0=0.05$, $0.15$, $...$, $0.25$. All the results are
obtained from MF calculations at $T=0.5$.

Fig.4 The temperature dependence of the loop area $A$ and the order
parameter $Q$ obtained from (a) MF calculations and (b) MC calculations,
with the AC field.

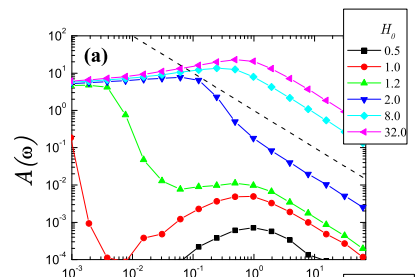
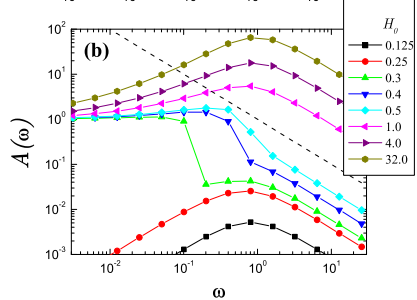
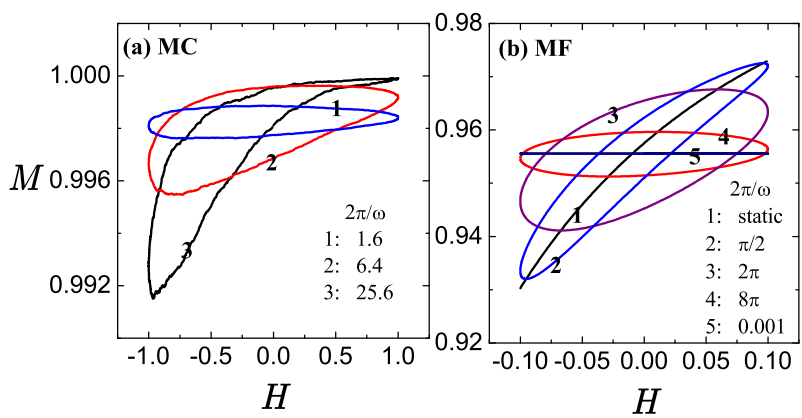
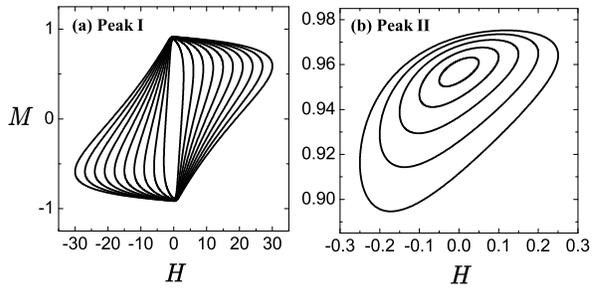
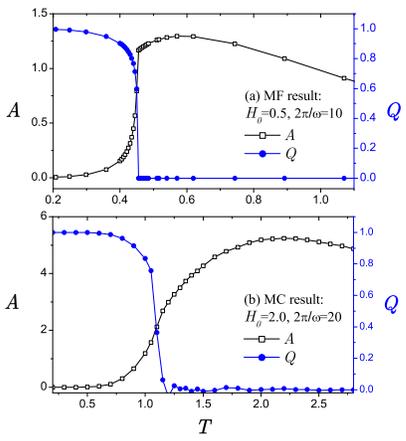
\end{document}